\documentclass[twocolumn,showpacs,preprintnumbers,amsmath,amssymb]{revtex4}
\usepackage{graphicx}
\begin{document}
\title
{
Orbital Magnetism 
in Three-Dimensional Quantum Dots
}

\author
{ 
Takeru {\sc Suzuki} 
,
Hiroshi {\sc Imamura}, 
Masahiko {\sc Hayashi} 
and Hiromichi {\sc Ebisawa}
}

\affiliation
{
Graduate School of Information Sciences, 
Tohoku University, Sendai 
980-8579\\
}

\begin{abstract}
We study orbital magnetism in 
a three-dimensional (3D) quantum dot 
with a parabolic confining potential.
We calculate the free energy of the system 
as a function of the magnetic field and 
the temperature. 
By this, we show that the temperature-field 
plane can be classified into three regions 
in terms of  
the characteristic behavior of 
the magnetization: 
the Landau diamagnetism, 
de Haas-van Alphen oscillation and 
mesoscopic fluctuation of magnetization. 
We also calculate numerically 
the magnetization of the system 
and then the current density distribution. 
As for the oscillation of 
the magnetization when the field is varied, 
the 3D quantum dot shows a longer period 
than a 2D quantum dot which   
contains the same number of electrons. 
A large paramagnetism appears at
low temperatures when 
the magnetic field is very weak. 
\end{abstract}

\maketitle
%
Orbital magnetism of a bulk system appears 
as the Landau diamagnetism (LD).
In the past, Landau's work has provoked 
discussion regarding  
his treatment of boundary~\cite{rf:1,rf:2}
and the effects of boundary 
on the orbital magnetism 
have been studied by many people.~\cite{rf:3,rf:4} 
Recently,  
two-dimensional (2D) quantum dots 
have become experimentally 
available with  the capability of 
controlling the dots' size, shape 
and number of electrons.~\cite{rf:5,rf:6,rf:7}
Stimulated by experimental development, 
many theorists have been studying on 
2D quantum dots with 
parabolic confining potentials. 
In particular, 
Ishikawa and Fukuyama (IF)~\cite{rf:3} 
have claimed that 
the temperature-field ($T$-$B$) 
plane can be classified into three regions 
in terms of  
the characteristic behavior of 
the magnetization $M(T,B)$: 
the LD, 
de Haas-van Alphen (dHvA) oscillation and 
mesoscopic fluctuation (MF) of magnetization. 
Such magnetic behavior is mainly governed by 
the energy levels near the Fermi level of 
the system. 
In the case of 3D quantum dots, 
the additional degree of freedom 
makes the density of states  
larger than 
that of 2D quantum dots. 
Hence, it is expected that 
the temperature affects the magnetization 
more easily, 
which changes
the magnetic behavior to some extent 
and 
possibly invalidates the classification. 

In this letter, we present our study on 
the orbital magnetism of 3D quantum dots. 
We consider 
noninteracting electrons 
confined in a parabolic potential 
and calculate analytically 
the free energy $\Omega$ of the system.
Thereby, we show that 
the temperature-field 
plane can still be classified into 
the above three regions 
in terms of 
the characteristic behavior of 
the magnetization, 
$M(T,B)=-\partial\Omega/\partial B$.   
In order to confirm this classification, 
we calculate numerically 
the magnetization of the system 
and then the current density distribution.  
In addition to the consistency 
between the analytical 
and numerical calculations, 
we report some remarkable 
findings of our numerical calculation. 
Hereafter, we set the fundamental constants as 
$\hbar=c=k_{\rm B}=1$. 
 
We consider spinless electrons 
of effective mass $m$ and charge $-e$ 
confined in a spherical quantum dot 
with a parabolic confining potential 
in a uniform magnetic field ${\mathbf B}$. 
The total Hamiltonian $H$ of the system
is given by 
the sum of single electron Hamiltonians as 
\begin{eqnarray}
    H
    &=&
    \sum_{i=1}^{N}
    \left \{
    \mbox{$\frac{1}{2m}$}
    \left[
    -{\rm i}{\mathbf \nabla}_i
    +
    e
    {\mathbf A}({\mathbf r}_i)
    \hspace{0.2em}
    \right]^2
    +
    \!\mbox{$\frac{m}{2}$}
    \omega_3^2 {\mathbf r}_i^2
    \right\}, 
    \label{eq:hamiltonian}
\end{eqnarray}
where $N$ is the total number of electrons, 
and 
${\mathbf A}$ is a vector potential 
satisfying 
${\mathbf B}={\rm rot}{\mathbf A}$. 
The last term represents 
the confining potential, where 
$\omega_3$ parametrizes the strength of 
the isotropic confinement. 

The eigenstates  
of the single electron Hamiltonian 
comprise  
the Fock-Darwin states in the $x$-$y$ plane 
~\cite{rf:3} 
and the eigenstates of a harmonic oscillator 
in the $z$-direction parallel to ${\mathbf B}$.
In the cylindrical coordinate $(\rho,\phi,z)$, 
the eigenfunction $\psi$ is given by 
\begin{equation}
    \psi_{n\alpha\nu}({\mathbf r})
    =
    \left(2\pi\right)^{-\frac{1}{2}}
    {\rm exp}({\rm i}\alpha\phi)
    \hspace{0.2em}
    R_{n\alpha}(\rho)
    \hspace{0.2em}Z_{\nu}(z), 
    \label{eq:wavefunction}
\end{equation}
where 
$n$, $\alpha$, $\nu$ are integers and 
$n$, $\nu \geq 0$.
$(n,\alpha,\nu)$ consists of 
a set of quantum numbers. 
$R_{n\alpha}(\rho)$ is 
the radial wave function of the Fock-Darwin 
state specified by $(n,\alpha)$  
and 
$Z_{\nu}(z)$ is  
the $\nu$-th eigenfunction of 
the harmonic oscillator.  
The eigenenergy $E_{n\alpha\nu}$ 
is given by  
\begin{eqnarray}
    &&
    E_{n\alpha\nu}
    =
   \left( 
    n+\!\mbox{$\frac{1}{2}$}
    \right)
    \omega_1
    +|\alpha|
    \hspace{0.2em}
    \omega _2
    +\left(
    \nu+\!\mbox{$\frac{1}{2}$}
    \right)
    \omega_3
    \label{eq:energylevel},  
    \\
    &&
    \hspace{4em}
    \omega_1
    =
    \sqrt{
    \omega_{\hspace{-0.1em}B}^2+4\omega_3^2
    },
    \label{eq:frequency_1}
    \\
    &&
    \hspace{4em}
    \omega_2(\pm)
    =
    \omega_\pm
    =
    \mbox{$\frac{1}{2}$}
    (\omega_1 \pm \omega_{\hspace{-0.1em}B}), 
    \label{eq:frequency_2}   
\end{eqnarray}
where $\omega_{\hspace{-0.1em}B}$ 
denotes   
the cyclotron frequency, 
$\omega_{\hspace{-0.1em}B}=eB/m$. 
In eq.~(\ref{eq:frequency_2}), 
the characteristic frequency 
$\omega_2=\omega_2(\sigma)$ 
depends on 
the sign $\sigma \hspace{0.2em}(=\pm)$ of 
the angular momentum $\alpha$.   
Hereafter, we set $\omega_3$ as the energy unit,   
though we keep 
the symbol $\omega_3$ in 
the several equations below for clarity. 

The magnetization $M$ 
of the entire system 
is given by 
$M=-(\partial\Omega / \partial B)_{T\mu}$,
~\cite{rf:8} 
where
$T$ is the temperature, 
$\mu$ is the chemical potential, 
and $\Omega$ is the free energy of the system, 
\(
\Omega=
-T\sum_{n\alpha \nu}
{\rm ln}
\{1+{\rm exp}
[(\mu-E_{n\alpha \nu})/T]
\}
\).
By means of 
the Poisson summation formula,~\cite{rf:9}  
the triple sum 
with respect to $n,\alpha,\nu$ is transformed 
into another triple sum, 
\( \Omega
    =
    \sum_{p,q,r=-\infty}^{\infty}
    \Omega_{pqr}, 
\)
where $\Omega_{pqr}$ is expressed as  
the Fourier integral which we can  
calculate analytically by assuming  
$\mu \gg T$ and $\mu \gg \omega_{\hspace{-0.1em}B}$. 
With these calculations carried out, 
the free energy $\Omega$ 
is given as follows, 
\begin{eqnarray}
    &&
    \Omega \approx 
    \Omega_0+\tilde{\Omega}, 
    \label{eq:Omega}
    \\
    &&
    \Omega_0
    =
    -\mbox{$\frac{1}{24}$}\mu^4
    +\left(
    \mbox{$\frac{1}{48}$}\omega_{\hspace{-0.1em}B}^2
    -\!\mbox{$\frac{\pi^2}{12}$} T^2
    +\!\mbox{$\frac{1}{16}$}
    \right)
    \mu^2,
    \label{eq:Omega0}
    \\
    &&
    \tilde{\Omega}
    =
    \sum_{i=1}^{3}
    \Omega_i  
    =
    \sum_{i=1}^{3}
    \sum_{\sigma=\pm}
    \sum_{p=1}^{\infty}
    \Omega^{\sigma}_{i}(p), 
    \label{eq:Omegaosc}
    \\
    &&
    \Omega^{\sigma}_{i}(p)
    =
    Q^{\sigma}_{i}(p)
    \hspace{0.2em}
    \Psi\hspace{-0.2em}
    \left(
    2p\pi^2T/\omega_i
    \right)
    {\rm cos}\hspace{-0.2em}
    \left(
    2p\pi\mu/\omega_i
    \right),
    \label{eq:Omegai}
    \\
    &&
    Q^{\sigma}_{i}(p)
    =
    -
    \frac
    {
    (-1)^{ip}\hspace{0.2em}
    \omega_i^3
    }
    {
    8\pi^4 \hspace{0.2em} p^4
    \omega_j \omega_k
    }
    \nonumber
    \\
    &&
    \hspace{1em}
    -
    \sum_{q=1}^{\infty}
    \frac{(-1)^{ip}}
    {4\pi^4 \hspace{0.1em} p^2}
    \left[ 
    \frac
    {
    (-1)^{jq}\hspace{0.2em}
    \omega_i^3 \omega_j
    }
    {
    \omega_k (p^2\omega_j^2-q^2\omega_i^2)
    }
    +
    \frac
    {
    (-1)^{kq}\hspace{0.2em}
    \omega_i^3 \omega_k
    }
    {
    \omega_j (p^2\omega_k^2-q^2\omega_i^2)
    }
    \right ]
    \nonumber
    \\
    &&
    \hspace{1em}
    -
    \sum_{q=1}^{\infty}
    \sum_{r=1}^{\infty}
    \frac
    {
    (-1)^{ip+jq+kr}\hspace{0.2em}
    \omega_i^3 \omega_j \omega_k
    }
    {2\pi^4 \hspace{0.1em}
    (p^2\omega_j^2-q^2\omega_i^2)
    (p^2\omega_k^2-r^2\omega_i^2)
    },
    \label{eq:Qi}
\end{eqnarray}
where 
$\Psi(x)=x/{\rm sinh}(x)$, 
and in eq.(\ref{eq:Qi}),  
the coefficients 
$Q^{\sigma}_{1},Q^{\sigma}_{2},Q^{\sigma}_{3}$ 
take 
the set of indices  
$(i,j,k)=(1,2,3),(2,3,1),(3,1,2)$, 
respectively.
The summation in eq.(\ref{eq:Qi}) 
is executed further and we obtain 
\begin{eqnarray}  
    Q^{\sigma}_{1}(p)
    &=&
    -
    \frac
    {(-1)^{p} 
    \hspace{0.2em}\omega_1
    }
    {
    8\pi^2\hspace{0.2em}
    p^2\hspace{0.2em}
    {\rm tan}\hspace{-0.2em}
    \left(  
    \pi p\omega_2/\omega_1
    \right)
    {\rm sin}\hspace{-0.2em}
    \left( 
    \pi p\omega_3/\omega_1
    \right)
    },
    \label{eq:Q1}
    \\
    Q^{\sigma}_{2}(p)
    &=&
    -
    \frac
    { 
    \hspace{0.2em}\omega_2
    }
    {
    8\pi^2\hspace{0.2em}
    p^2\hspace{0.2em}
    {\rm sin}\hspace{-0.2em}
    \left( 
    \pi p\omega_3/\omega_2
    \right)
    {\rm sin}\hspace{-0.2em}
    \left( 
    \pi p\omega_1/\omega_2
    \right)
    },
    \label{eq:Q2}
    \\
    Q^{\sigma}_{3}(p)
    &=&
    -
    \frac
    {(-1)^{p}
    \hspace{0.2em}\omega_3
    }
    {
    8\pi^2\hspace{0.2em}
    p^2\hspace{0.2em}
    {\rm sin}\hspace{-0.2em}
    \left( 
    \pi p\omega_1/\omega_3
    \right)
    {\rm tan}\hspace{-0.2em}
    \left( 
    \pi p\omega_2/\omega_3
    \right)
    }.
    \label{eq:Q3}
\end{eqnarray}

The free energy $\Omega_0$ 
in eq.~(\ref{eq:Omega0}) 
is the nonoscillating part 
with respect to the magnetic field,   
while 
$\tilde{\Omega}=\Omega_1+\Omega_2+\Omega_3$  
in eq.~(\ref{eq:Omegaosc}) 
represents 
the oscillating part of  
the total free energy $\Omega$. 
For the oscillating part, 
the free energy  
$\Omega_1=\sum_\sigma \sum_p \Omega^{\sigma}_{1}(p)$ 
vanishes because 
\(
\sum_{\sigma}\Omega^{\sigma}_{1}(p)
\propto
\sum_{\sigma}Q^{\sigma}_{1}(p)
\)
is identically zero 
due to eq.~(\ref{eq:frequency_2}). 
As for $\Omega_2$,  
$\Omega^{\pm}_{2}(p)$
has the oscillating factor, 
${\rm cos}(2p \pi \mu/\omega_\pm)$, 
in eq.~(\ref{eq:Omegai}) 
as a function of the field parameter $\omega_{\hspace{-0.1em}B}$   
since $\omega_\pm$ varies monotonically 
with respect to $\omega_{\hspace{-0.1em}B}$. 
Although the remaining factor $Q^{\pm}_{2}(p)$ 
in $\Omega^{\pm}_{2}(p)$ 
also oscillates by 
${\rm sin}(\pi p \omega_3/\omega_\pm)
{\rm sin}(\pi p \omega_1/\omega_\pm)$ 
in eq.~(\ref{eq:Q2}), 
we can neglect this oscillation,  
compared with ${\rm cos}(2p \pi \mu/\omega_\pm)$,  
since we can assume 
$\omega_3 \ll \mu$ 
for the sufficiently large size of the system
and also 
$\omega_1 \ll \mu$ 
due to 
the assumption, $\mu \gg \omega_{\hspace{-0.1em}B}$.  
As for $\Omega_3$, 
$\Omega^{\sigma}_{3}(p)$ 
in eq.~(\ref{eq:Omegai}) has only
a negligible oscillation 
caused by 
${\rm sin}(\pi p \omega_1/\omega_3)
{\rm tan}(\pi p \omega_2/\omega_3)$
in eq.~(\ref{eq:Q3}), 
because ${\rm cos}(2p \pi \mu/\omega_3)$ 
is independent of the magnetic field. 
Consequently,  
$\Omega_{2}$ is dominant 
in eq.~(\ref{eq:Omegaosc}) in terms of 
the oscillation of magnetization. 
Hereafter, we write the effective part of 
the free energy $\tilde{\Omega}$ as 
\begin{eqnarray}
  \tilde{\Omega}_{\rm eff}
  =
  \sum_{\sigma=\pm}
  \sum_{p=1}^{\infty}
  Q^{\sigma}_{2}(p)
    \Psi\hspace{-0.2em}
    \left(
    2p\pi^2T/\omega_\sigma 
    \right)
    {\rm cos}\hspace{-0.2em}
    \left(
    2p\pi\mu/\omega_\sigma
    \right), 
    \hspace{2em}
  \label{eq:effective}
\end{eqnarray}
where $\omega_\sigma=\omega_2(\sigma)$. 

It should be noted, however, that  
eqs.~(\ref{eq:Q1})-(\ref{eq:Q3}) have
spurious divergent singularities 
when $\omega_i/\omega_j$ is a rational number. 
To eliminate such divergences, 
we rewrite 
$\tilde{\Omega}$ as defined in 
eqs.~(\ref{eq:Omegaosc})-(\ref{eq:Qi}),  
by rearranging terms, into 
\begin{eqnarray}
    &&
    \tilde{\Omega}
    =
    \sum_{\sigma=\pm}
    \left(
    \sum_{p=1}^{\infty}
    \hat{\Omega}_p^\sigma
    +
    \sum_{p,q=1}^{\infty}
    \hat{\Omega}_{p,q}^\sigma
    +
    \sum_{p,q,r=1}^{\infty}
    \hat{\Omega}_{p,q,r}^\sigma
    \right),
    \label{eq:rewritten}
    \\
    &&
    \hat{\Omega}_p^\sigma
    =
    -
    \frac{
    (8 \pi^4)^{-1}
    }
    {\omega_1\omega_2\omega_3}
    \sum_{i=1}^3
    (-1)^{ip}
    \frac{
    Y
    \left(
    p/\omega_i
    \right)
    }
    {(p/\omega_i)^4},
    \hspace{2em}
    \label{eq:Omega_p}   
    \\
    &&
    \hat{\Omega}_{p,q}^\sigma
    =
     -
    \frac{
    (4 \pi^4)^{-1}
    }
    {\omega_1\omega_2\omega_3}
    \sum_{(i,j)}
    (-1)^{ip+jq}
    \frac
    {
    \frac
    {Y\left( p/\omega_i \right)}
    {(p/\omega_i)^2}
    -
    \frac
    {Y\left( q/\omega_j \right)}
    {(q/\omega_j)^2}
    }
    {(p/\omega_i)^2 - (q/\omega_j)^2},
    \hspace{2em}
    \label{eq:Omega_pq}   
    \\
    &&
    \hat{\Omega}_{p,q,r}^\sigma
    =
    -
    \frac{
    (2 \pi^4)^{-1}
    }
    {\omega_1\omega_2\omega_3}
    (-1)^{p+r}
    \nonumber
    \\
    && 
    \phantom{ \hat{\Omega}_{p,q,r}^\sigma}
    \times
    \frac
    {
    \frac
    {Y\left( p/\omega_1 \right)
    -Y\left( r/\omega_3 \right)}
    {(p/\omega_1)^2 - (r/\omega_3)^2}
    -
   \frac
    {Y\left( q/\omega_2 \right)
    -Y\left( r/\omega_3 \right)}
    {(q/\omega_2)^2 - (r/\omega_3)^2}
    }
    {(p/\omega_1)^2 - (q/\omega_2)^2}
    ,
    \hspace{2em}
    \label{eq:Omega_pqr}
\end{eqnarray}
where 
\(
Y(x)=
\Psi(2\pi^2 T x)
\hspace{0.2em}
{\rm cos}(2\pi \mu x)
\), 
and in eq.(\ref{eq:Omega_pq}),  
the index $(i,j)$ takes  
$(1,2),(2,3)$ and $(3,1)$. 
Equations~(\ref{eq:Omega_p})-(\ref{eq:Omega_pqr}) 
are apparently 
free from divergences. 
It is expected that 
as the temperature $T$ rises, 
the total amplitude of oscillation 
is rapidly suppressed by 
the exponentially decreasing factor, 
$\Psi(2p\pi^2 T/\omega_i)\sim
(4p\pi^2T/\omega_i) \hspace{0.2em}
{\rm exp}(-2p\pi T/\omega_i)$. 

At sufficiently high temperatures, 
it is expected that 
$\Omega_0$ in eq.~(\ref{eq:Omega0}) 
makes a dominant contribution 
to the total free energy $\Omega$,
and
the magnetization $M_0$ is given by 
\begin{equation}
    M_0/\mu_{\rm B}
    = 
    -
    2
    \left(
    \partial\Omega_0/
    \partial \omega_{\hspace{-0.1em}B}
    \right)_{T\mu}
    = 
    -\mbox{$\frac{1}{12}$}
    \mu^2\omega_{\hspace{-0.1em}B}, 
    \label{eq:Landau}
\end{equation}
where 
$\mu_{\rm B}=e/(2m)$ is the Bohr magneton. 
The magnetization $M_0$ is negative and 
its absolute value increases linearly  
with increasing $\omega_{\hspace{-0.1em}B}$,   
which is characteristic of  
the LD  
in a bulk system.~\cite{rf:9}
%
On the other hand, at low temperatures, 
by using the free energy 
$\tilde{\Omega}_{\rm eff}$ 
in eq.~(\ref{eq:effective}), 
we obtain an oscillating part of 
the magnetization,  
\(
\tilde{M}/\mu_{\rm B}=-2
(\partial\tilde{\Omega}_{\rm eff}/
\partial \omega_{\hspace{-0.1em}B})_{T\mu}
\), 
which contains the trigonometric functions of 
$2\pi p\mu/\omega_+$ and  
$2\pi p\mu/\omega_-$. 
It should be noted that 
as the temperature $T$ decreases, 
the amplitude of the oscillation of $\tilde{M}$ 
becomes larger. 
%

\begin{figure}
  \includegraphics[width=0.8\columnwidth]{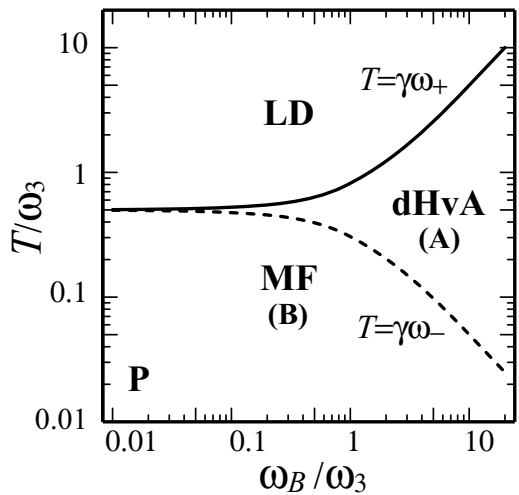}
\vspace{-10pt}
\caption{
Temperature-field ($T$-$B$)  
plane can be classified into three regions: 
the Landau diamagnetism (LD), 
the de Haas-van Alphen (dHvA) oscillation, 
and the mesoscopic fluctuation (MF), 
in terms of the characteristic behavior of 
the magnetization $M(T,B)$. 
$\omega_{\hspace{-0.1em}B}$ is the cyclotron frequency 
and $\omega_3$ parametrizes the strength 
of the confinement. 
In the region indicated by P,  
the remarkable paramagnetism appears. 
}
\label{fig:1}
\vspace{-5pt}
\end{figure}
%
In the following, we discuss how the three regions 
introduced by IF~\cite{rf:3} are 
modified in 3D. 
We regard the LD region as the region 
where 
the amplitudes of the oscillations 
specified by $p$ 
are much smaller than  
the confining energy $\hbar\omega_3$, 
which is unity 
in our notation.  
Hence, the criterion for the LD region is 
given by
\(
|Q_{2}^{\pm}(p)|
\Psi(2p\pi^2T/\omega_\pm)\ll 1 
\)
 for all $p$.
Introducing the cutoff 
$\epsilon \hspace{0.2em}(\ll 1)$, 
we defined the LD region 
as the region where 
the following two inequalities 
hold simultaneously, 
\begin{equation}
    \Psi\hspace{-0.2em}
    \left(2p\pi^2T/\omega_+\right) 
    \leq \epsilon,
    \hspace{1em}
    \Psi\hspace{-0.2em}
    \left(2p\pi^2T/\omega_-\right)
    \leq \epsilon.   
    \label{eq:cut-off}
\end{equation}
In this letter, we set $\epsilon=0.1\%$, but 
it will be necessary, however, to reexamine 
the value of $\epsilon$ 
when our analytic classification  
is compared to any  
experiment. 
Noting 
$\Psi'(x)\hspace{-0.2em}
<\hspace{-0.2em}0$ and 
\(
\omega_{+}
\hspace{-0.3em}
>\hspace{-0.2em}\omega_{-}\), 
the region determined by eq.~(\ref{eq:cut-off}) 
is equivalent to 
$T \geq \gamma \omega_+$, 
where 
$\gamma=\Psi^{-1}(\epsilon)/(2\pi^2)\approx 0.5$. 
As a result,  
the boundary between 
the LD and the oscillatory region is 
given by 
$T=0.5\hspace{0.2em}\omega_+$, as plotted 
by the solid line in Fig.1. 

The oscillatory region $T<\gamma\omega_+$  
is divided into two subregions 
as shown in Fig.1; 
(A) $\gamma\omega_{+}>T\geq \gamma\omega_{-}$, 
and 
(B) $\gamma\omega_{-}>T$.
In subregion (A), the first inequality 
in eq.~(\ref{eq:cut-off}) fails  
while the second one still holds.
Hence, only  
$Q_2^{+}(p)\Psi(2\pi^2T/\omega_+)
{\rm cos}(2\pi \mu/\omega_+)$  
contributes to $\tilde{\Omega}_{\rm eff}$.   
The effective free energy 
$\tilde{\Omega}^{\rm A}_{\rm eff}$
is given by 
\(
\tilde{\Omega}^{\rm A}_{\rm eff}
= \sum_{p=1}^{\infty}
Q_2^{+}(p)
\Psi\hspace{-0.2em}
\left(
2p\pi^2T/\omega_+
\right)
{\rm cos}\hspace{-0.2em}
\left(
2p\pi\mu/\omega_+
\right)
\).
On the other hand, 
both inequalities fail in subregion (B), 
where 
two types of oscillations regarding 
$\omega_\pm$ contribute to 
$\tilde{\Omega}_{\rm eff}$, 
and 
the free energy 
$\tilde{\Omega}^{\rm B}_{\rm eff}$  
is given by eq.~(\ref{eq:effective}). 
As the temperature $T$ is lowered, 
the terms with larger $p$ start 
contributing to 
$\tilde{\Omega}_{\rm eff}$. 

For subregion (A), 
magnetization $\tilde{M}$ 
derived from $\tilde{\Omega}^{\rm A}_{\rm eff}$
contains a trigonometric function  
having the period $\omega_+/\mu$. 
The dHvA region 
corresponds to the region 
$\gamma\omega_- \leq T < \gamma\omega_+$, 
characterized by the oscillation 
having the period $\omega_+/\mu$. 
Particularly in a strong field region, 
since $\omega_+ \sim \omega_{\hspace{-0.1em}B}$,
magnetization $\tilde{M}$ 
behaves as the periodic function 
of variable $\omega_{\hspace{-0.1em}B}^{-1}$. 
This periodicity is typical of 
the dHvA oscillation 
in a bulk system.~\cite{rf:8,rf:9}

For subregion (B), 
magnetization $\tilde{M}$ 
derived from $\tilde{\Omega}^{\rm B}_{\rm eff}$
contains 
a trigonometric function 
having the period $\omega_-/\mu$ 
in addition to $\omega_+/\mu$.
In this region, 
$\omega_+$ and $\omega_-$ are not so close to 
each other and the two modes fluctuate separately. 
Consequently, 
the coexistence of the above two modes 
causes the rapid and irregular oscillation 
of magnetization.  
This magnetic behavior 
corresponds to the MF. 
Hence, subregion (B) is regarded as the MF region.  

However, 
when the magnetic field is extremely weak 
in the MF region, 
the periods of $\omega_\pm/\mu$
become much closer since 
$\omega_\pm \sim 1\pm \frac{1}{2}\omega_{\hspace{-0.1em}B}$. 
It is impossible to regard
$\omega_\pm$ as distinguishable 
and therefore   
the separation of the two fluctuations
breaks down in the above argument. 
In this case, the largest contribution   
to $\tilde{\Omega}$ 
in eq.~(\ref{eq:rewritten}) comes from 
the modes satisfying 
$p:q:r=\omega_1:\omega_2:\omega_3$, 
namely,   
\(
p/\omega_1-q/\omega_2=
q/\omega_2-r/\omega_3=
r/\omega_3-p/\omega_1=0
\) 
for $\hat{\Omega}_{p,q,r}^{\sigma}$
in eq.~(\ref{eq:Omega_pqr}). 
From the behavior of
characteristic frequency  
$\omega_1 \sim 2 + \frac{1}{4}\omega_{\hspace{-0.1em}B}^2$, 
we can see that 
the modes of $p=2j$, $q=j$, $r=j$ 
($j$ is a natural number) 
make a dominant contribution to
magnetization $\tilde{M}$. 
The largest contribution to $\tilde{M}$ 
is evaluated as 
\begin{eqnarray}
   \frac{\tilde{M}}{\mu_{\rm B}}
   \sim
   \sum_{j=1}^\infty
   \frac{
   \hat{M}
   _{2j,j,j}}
   {\mu_{\rm B}}
   =
   -2
   \sum_{\sigma=\pm}
   \sum_{j=1}^\infty
   \left(
   \frac
   {\partial
   \hat{\Omega}
   _{2j,j,j}
   ^{\sigma}}
   {\partial \omega_{\hspace{-0.1em}B}}
   \right)_{T\mu}. 
   \hspace{1em}
   \label{eq:mode211}
\end{eqnarray}
Furthermore, we can estimate 
the upper bound $U(p,q,r)$ 
of $|\hat{\Omega}^{\sigma}_{p,q,r}|$
by using the triangle inequality and 
$|Y(x)| \leq 1$ for eq.(\ref{eq:Omega_pqr}), 
and it is found that 
the upper bound $U(2j,j,j)$ decreases as 
$U(2j,j,j)=j^{-4}U(2,1,1)$ for larger $j$. 
Hence, it is expected that a few modes 
with small $j$ 
make a major contribution.  
The mode with $j=1$,   
$\hat{M}_{2,1,1}$,  
is plotted in Fig.2(a), 
where we take $T=0.01$ and $N=5000$. 
The magnetization curve 
in Fig.2(a) 
shows a remarkable paramagnetism 
at weak magnetic fields 
below $\omega_{\hspace{-0.1em}B} \sim 0.02$. 
The region corresponding to 
the paramagnetism 
is indicated by P in Fig.1.

Next, we describe the procedure of 
the numerical calculation for magnetization. 
We consider magnetization as 
the total dipole moment of electrons with the 
confining potential. 
From this standpoint,
magnetization ${\mathbf M}$ is given by 
\(
    {\mathbf M}
    =
    \int_{\infty}
    {\rm d}V
    \mbox{$\frac{1}{2}$}
    \left[
    {\mathbf r}
    \times 
    {\mathbf J}({\mathbf r})
    \right]
\), 
where ${\mathbf J}({\mathbf r})$ is the 
current density distribution of the system.
The calculation of $\mathbf{M}$ reduces to 
that of ${\mathbf J}({\mathbf r})$ 
which is given by  
\(
    {\mathbf J}({\mathbf r})
    =
    {\rm Re}
    \sum_{s=(n,\alpha,\nu)}
    f(E_s)
    \psi^\ast_s({\mathbf r})
    \frac{-e}{m}
    \left(
    {\rm -i}
    {\mathbf \nabla}+
    e
    {\mathbf A}
    \right)
    \psi_s({\mathbf r})
\).  
Here,  
$f(E)$ denotes  
the Fermi distribution function,  
$f(E)=\{ {\rm exp}[(E-\mu)/T]+1 \}^{-1}$.  
By using eq.(\ref{eq:wavefunction}), 
the current density distribution 
leads to 
${\mathbf J}({\mathbf r})=J(\rho,z){\mathbf e}_\phi$,  
and 
\begin{eqnarray}
    J(\rho,z)=
    - \sqrt{\frac{e}{m}}
    \hspace{-2em}
    \sum_{\hspace{2em}s=(n,\alpha,\nu)}
    \hspace{-2em}
    f(E_s)
    \left|
    \psi_{s}({\mathbf r})
    \right|^2
    \left(
    \alpha \frac{\xi}{\rho}
    +
    \frac{\omega_{\hspace{-0.1em}B}}{2}
    \frac{\rho}{\xi}
    \right),
    \hspace{2em}
    \label{eq:current}  
\end{eqnarray}
where 
$\xi= m^{-\frac{1}{2}}$. 
By using eq.(\ref{eq:current}), 
the total dipole moment $\mathbf{M}$ is
calculated analytically~\cite{rf:3} 
and then 
${\mathbf M}=M{\mathbf e}_z$. 
The $z$-component $M$ is given by 
\begin{eqnarray}
    M/
    \mu_{\rm B}
    =
    -
    \hspace{-2em}
    \sum_{\hspace{2em}s=(n,\alpha,\nu)}
    \hspace{-2em}
    f(E_s)
    \left[
    \alpha
    +\left(
    2n+|\alpha|+1
    \right)
    \hspace{0.2em}
    \omega_{\hspace{-0.1em}B}/\omega_1
    \right],
    \hspace{2em}
    \label{eq:magne}  
\end{eqnarray}
which is also derived from 
\(
M/\mu_{\rm B}=
-2(\partial \Omega/\partial \omega_{\hspace{-0.1em}B})_{T\mu}
=
-2\sum_s f(E_s)
(\partial E_s/\partial \omega_{\hspace{-0.1em}B})
\) 
due to eq.(\ref{eq:energylevel}). 
This coincidence of two derivations, 
of course, 
occurs regardless 
of the type of confining potentials. 
In the numerical calculation,  
we calculate the chemical potential first  
by solving $ N=\sum_s f(E_s)$ for $\mu$, 
and then compute magnetization $M$ 
by using eq.(\ref{eq:magne}). 

\begin{figure}
  \includegraphics[width=1.0\columnwidth]{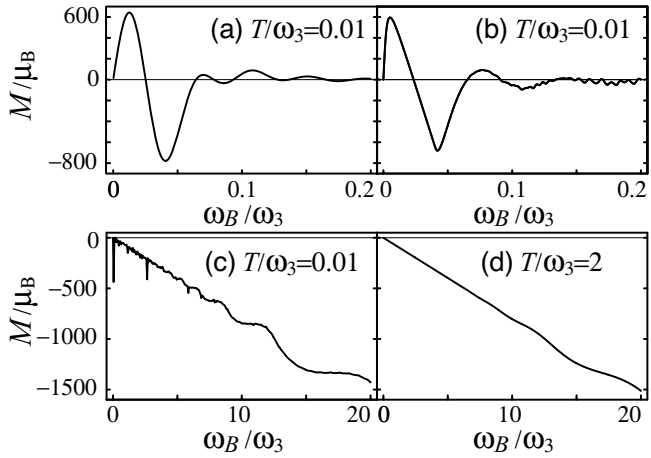}
\vspace{-10pt}
\caption{
Figure 2(a) shows the contribution 
$\hat{M}_{2,1,1}$,   
which is the lowest mode $(j=1)$ 
in eq.~(\ref{eq:mode211}).
Figures 2(b)-2(d) show 
the magnetization curve obtained 
by the numerical calculation for $N=5000$. 
}
\label{fig:2}
\vspace{-5pt}
\end{figure}
%
Magnetization $M$ 
calculated numerically 
is shown in Fig.2 
for $N=5000$. 
At low temperatures 
($T=0.01$, Figs.2(b)
and 2(c)), 
the magnetization oscillates 
over the entire range of magnetic fields.
At extremely weak fields  
below $\omega_{\hspace{-0.1em}B} \sim 0.1$
(Fig.2(b)), 
the magnetization oscillates smoothly 
with significantly large amplitude. 
The smoothness implies 
that not so many modes are dominant 
in the oscillation. 
Moreover, Fig.2(b) 
agrees qualitatively with 
the behavior of $\hat{M}_{2,1,1}(T,\mu,B)$ 
(Fig.2(a)) with 
the chemical potential 
$\mu(T,N,B)$ 
calculated numerically 
for $T=0.01$ and $N=5000$. 
The large smooth oscillation 
in Fig.2(b) 
is consistent 
with our analysis in terms of free energy. 
At weak magnetic fields 
above $\omega_{\hspace{-0.1em}B}\sim 0.1$, 
the large oscillation is suppressed and 
the magnetization irregularly takes 
a positive or negative value. 
This fluctuation continues 
until $\omega_{\hspace{-0.1em}B}\sim 5$ 
with the linear magnetization superposed 
(Fig.2(c)).
At high magnetic fields 
above $\omega_{\hspace{-0.1em}B}\sim 5$, 
the large smooth oscillation is 
clear in Fig.2(c). 
It is considered that 
this magnetization curve
reflects the dHvA oscillation. 
On the other hand, 
at high temperatures 
($T=2$, Fig.2(d)), 
the magnetization is linear in $\omega_{\hspace{-0.1em}B}$
until $\omega_{\hspace{-0.1em}B}\sim 10$ and then  
the dHvA oscillation increases gradually. 

Compared with the magnetization in 2D  
reported by IF,~\cite{rf:3}
it is found that 
the period of oscillation 
is longer in 3D. 
For the 2D system, they also 
calculated the free energy analytically 
and 
its oscillating part is expressed as 
\(
\tilde{\Omega}=
\sum_{i=1}^{2}
\sum_\sigma\sum_p \Omega_i^\sigma (p)
\), and 
\(
\Omega_i^\sigma (p)
=
Q_i^\sigma (p)
\Psi(2p\pi^2 T/\omega_i)
\hspace{0.2em}
{\rm sin}(2p\pi\mu/\omega_i)
\),
in our notation. 
Here, we find that 
for 2D and 3D,   
these oscillating factors 
have the common phase $2p\pi\mu/\omega_i$. 
Noting the common phase, 
when $\omega_{\hspace{-0.1em}B}$ is varied, 
the period with respect to $\omega_i^{-1}$, 
$\Delta(1/\omega_i)$, 
is given by 
$\Delta(1/\omega_i)=p^{-1}\mu^{-1}$.  
Hence, $\mu^{-1}$ serves as 
the scale for the period of oscillation 
and the difference between the periods can be  
explained by 
that between the chemical potentials. 
The chemical potentials 
$\mu_{\rm 3D},\mu_{\rm 2D}$ 
are estimated by 
$N\approx -(\partial \Omega_0/\partial \mu)_{TB}$ 
and then  
\(
\mu_{\rm 3D} \approx \sqrt[3]{6N}
< 
\mu_{\rm 2D} \approx \sqrt{2N}
\) for the same number of electrons. 
Consequently, 
the period of oscillation in 3D is 
longer than that in 2D. 

\begin{figure}[]
  \includegraphics[width=1.0\columnwidth]{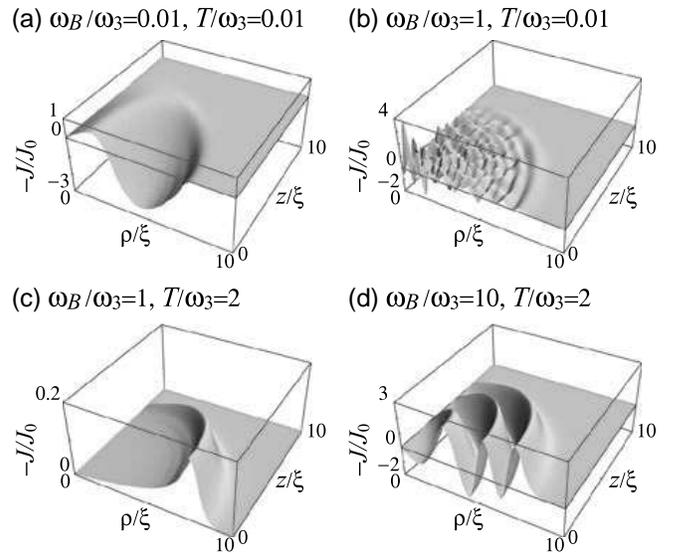}
\vspace{-10pt}
\caption{
Figures 3(a)-3(d) show 
the current density distribution 
$J(\rho,z)$ for $N=5000$ 
obtained by the numerical calculation. 
$\rho$ is radius and $z$ is height parallel to 
the magnetic field ${\mathbf B}$. 
Here, $\xi=1/\sqrt{m \omega_3}$ and 
$J_0=e\omega_3/(4\pi\xi^2)$. 
Four figures correspond to the 
characteristic regions 
of magnetization.
}
\label{fig:3}
\vspace{-5pt}
\end{figure}
%
The current density distribution 
$J(\rho,z)$ is 
computed by using eq.(\ref{eq:current}) and
is shown in Fig.3 
for $N=5000$. 
The positive value of the vertical axis 
corresponds to diamagnetism. 
When the remarkable paramagnetism 
appears (Fig.3(a)), 
a paramagnetic region 
for $J(\rho,z)$ extends 
throughout the spherical quantum dot 
except for a small region 
near the $z$-axis ($\rho=0$). 
In the MF region (Fig.3(b)), 
the paramagnetic and diamagnetic regions
are distributed irregularly. 
In the LD region (Fig.3(c)), 
the current density distribution
is localized near the surface of the quantum dot.
This surface current corresponds to
the edge current obtained 
by IF~\cite{rf:3} for the 2D system. 
In the dHvA region (Fig.3(d)), 
the current density distribution
oscillates smoothly with a large amplitude. 

In conclusion, we have studied 
orbital magnetism in a 3D quantum dot  
with a parabolic confining potential. 
From the analysis of the free energy, 
we have determined the boundaries 
between the LD, dHvA, and MF regions. 
We have also computed the magnetization 
of the system and then 
the current density distribution, 
which indeed reflects the character of 
each region. 
The results of numerical calculation show 
a large paramagnetism when the magnetic fields 
are extremely weak in the MF region. 
The dHvA oscillation has a longer period in 3D 
than that in a 2D quantum dot. 
These results are explained 
in terms of the free energy.  
Furthermore, we have found
surface current in the LD region, 
corresponding to
edge current for 2D quantum dots. 

The authors are grateful to H. Fukuyama and 
T. Sasaki for their useful discussions.
H. I. is supported by MEXT, Grant-in-Aid for 
Encouragement of Young Scientists, 13740197.


%
%

\begin{thebibliography}{99}
\bibitem{rf:4} 
The theoretical studies 
on effects of boundary 
are listed in the reference of 
the above paper.~\cite{rf:3}
\end{thebibliography}
\end{document}